# Dark Matter Redistribution Explains Galaxy Growth and Rotation Curve Development

Daniel E. Friedmann
2578 West 7 Avenue, Vancouver, BC Canada V6K 1Y9
friedmann.daniel@alumni.ubc.ca

**ABSTRACT**

There are significant discrepancies between observational evidence and the hierarchical galaxy formation theory with respect to the shape of dark matter halos, the correlation between galaxy characteristics, and galaxy evolutionary history. This paper introduces a modification to the hierarchical galaxy formation theory that hypothesizes that dark matter enters into highly elliptical orbits, and is therefore, effectively redistributed during the period of galactic nuclei activity. Adding this modification, the theory more accurately predicts the observed development history of galaxies and their resulting mature state. In particular, this modification predicts that galaxies grow in size, but not in mass, at an early time (~7 to 10 billion years ago), and develop their characteristic rotation curves which, at large radius, exhibit a relatively flat shape versus the expected Keplerian decline.

## 1. Introduction

The presently favored theory of galaxy formation, within the framework of Λ cold dark matter (CDM) cosmology, hypothesizes that small quantum fluctuations lead to dark matter 'halos'. Subsequently, halos cluster and merge into larger halos while simultaneously collecting gas. These structures form the seeds of galaxies that grow via further merging. This hierarchical galaxy formation theory (Davis *et al* 1985; White and Frenk 1991; Navarro *et al* 1995, 1996; Baugh 2006) has made several successful predictions of how the universe developed from a smooth initial state, as observed in the cosmic microwave background radiation, to its current large scale structure, and in particular, the three dimensional morphology of the galaxy distribution. However, observational evidence challenges the accuracy of critical aspects of our current understanding of the formation of galaxies (Bahcall and Casertano 1985; Persic *et al* 1999; Battaner and Florido 2000; Sancisi 2003; Glazebrook *et al* 2004; Daddi *et al* 2005; Trujillo *et al* 2006, 2007; Buitrago *et al* 2008; Disney *et al* 2008; Collins *et al* 2009; Garcia-Appadoo *et al* 2009; vanDokkum *et al* 2009; Cassata *et al* 2010; Williams *et al* 2010).

The first key disagreement between observation and theory relates to the shape of the inner cores of CDM halos. N-body simulations of the hierarchical galaxy formation theory predict the shape of dark matter halos. The galaxy rotation curves, which indicate the enclosed mass as a function of galactic radius, imply the observed shape of dark matter halos. Galactic rotation curves are plots of the rotational velocity of objects or gas orbiting the galactic centre versus their radius of rotation. These curves have a universal shape (Persic *et al* 1999) that is flat for large distances beyond the majority of visible matter. Whereas, simulations of the merging of dark matter halos indicate that they develop into cusp shaped NFW halos (Navarro *et al* 1996), observations of galaxy rotation curves reveal that halos are core shaped (Navarro and Steinmetz 2000; Battaner and Florido 2000; Donato *et al* 2009; Salucci and Martins 2009)

The next challenge to the theory comes from the systematic measurement of galaxy characteristics. If galaxies are indeed a result of a random hierarchical merger process, then their basic parameters should be independent. However, observation has revealed that there is a very strong correlation between six independent parameters (Disney *et al* 2008; Garcia-Appadoo *et al* 2009), implying that galaxy formation is controlled by a single parameter yet to be identified. Furthermore, rotation curve studies have revealed that the visible matter and dark matter distribution in galaxies is linked, with visible matter dominating in the inner (disk) region of galaxies and dark matter dominating in the outer (halo) region of galaxies (Sancisi 2003), indicating further evidence of a strong correlation between the key galaxy matter constituents. In particular, galaxy rotation curves for high surface brightness spiral galaxies reveal a featureless flat transition between the disk and halo regions. This unexplained observation is known as the 'halo-disk conspiracy' (Bahcall and Casertano 1985; Battaner and Florido 2000). In other words, rather than the theoretical prediction of a haphazard structure, the observational evidence indicates a structure with a high degree of organization.

Another challenge to the hierarchical galaxy formation theory has recently emerged, as observations of the very first galaxies have become available. The hierarchical formation process predicts an evolutionary history of galaxies, and in particular, that the most massive objects form last. Near infrared observations have revealed that a significant fraction of the most massive galaxies are already abundant 3 to 6 billion years (BY) after the Big Bang (Glazebrook *et al* 2004; Williams *et al* 2010), and are almost fully assembled to 90 per cent of their stellar mass versus the theoretically predicted 22 per cent (Collins *et al* 2009). Although the early assembly of massive galaxies maybe explainable via cold gas accretion in massive halos (Kang *et al* 2010), the hierarchical formation process also predicts that galaxy mass and size grow together; however, observations may indicate a different result. It has been found that the oldest most luminous galaxies in the early universe (at $z \sim 2.7$) are surprisingly compact, with an effective radius of approximately 1 kilo parsec (kps) or less in size (Daddi *et al* 2005; Trujillo *et al* 2006, 2007; Buitrago *et al* 2008; vanDokkum *et al* 2009; Cassata *et al* 2010). The observation that older spheroid galaxies have stellar masses similar to that of present day galaxies leads to the conclusion that they have grown in size by a factor of $\sim 4$ to 6, with a less significant growth in mass. Observations of galaxies, at a given stellar mass, at different look back times show that spheroid-like objects were smaller by a factor of 4.3, at $z \sim 2.7$; however, they show that disk-like objects were only a factor of 2.6 smaller (Trujillo *et al* 2007; Buitrago *et al* 2008). In addition, the massive Hubble sequence as a whole at z of 2.3 is very different from that in the local universe (Kriek *et al* 2009). Quiescent galaxies are much more compact and galaxies with a high star formation rate have irregular and clumpy structures that do not resemble today's spiral galaxies. Massive irregular galaxies with similar high star formation rates are very rare in the local universe and compact highly massive galaxies are nonexistent today.

The evolutionary picture that emerges is that galaxies experience an early period of rapid growth in mass, rather than a prolonged hierarchical assembly (Collins *et al* 2009). This early period is followed by a period of significant growth in galaxy size (with relatively smaller growth in mass, if any) and change in galaxy shape, which results in galaxies, by about 7 BY ago, that are not significantly different (in Hubble sequence and shape) to typical galaxies nearby.

Various models for the evolution of galaxy size and mass with time have been proposed and tested (Hopkins *et al* 2009; vanDokkum *et al* 2009). Hopkins *et al* conclude that late dry mergers are the only model that can explain the different observations for massive spheroid-like galaxies. However, there can only be enough mergers to account for about half the observed galaxy growth. The authors have to invoke a combination of all the other models, in addition to late dry mergers, to achieve a full accounting of the galaxy growth. These models do not deal with the difference in the growth factors of spheroid-like objects and disk-like objects. Furthermore, these models do not explain the first two challenges discussed above. Finally, the models predict a gradual evolution in the size of galaxies with time and a small decrease in stellar velocity dispersion with galaxy growth. The current observational evidence is still unclear on whether galaxies evolve slowly or fast (for the period $z > 1.5$) and whether the velocity dispersion decreases by a small or large factor as galaxies grow (Cenarro and Trujillo 2009; vanDokkum *et al* 2009, 2010).

The discrepancies between observations and theoretical predictions, and between the growth factor of spheroid-like objects and disk-like objects, suggest that a modification of the hierarchical galaxy formation theory is required. In addition, combining the late dry merger model (Hopkins *et al* 2009) for the period post $z = 1.5$, with another galaxy growth model for the period at larger z, opens the possibility for different theoretical predictions of key observations relating to velocity dispersion in early galaxies and rate of galaxy size evolution.

This paper explores a modification of the hierarchical galaxy formation theory that affects the evolutionary history of galaxies and their resulting visible matter and dark matter distribution. The modification hypothesizes that redistribution of dark matter occurs during the active galactic nuclei period of the galaxy. First, I discuss how such redistribution of dark matter might occur. Second, I explore how the modification to the theory affects the distribution of dark matter, the extent of the visible matter, and the evolution of galaxies. A new picture emerges with a stronger alignment between theory and observation.

**2. The Hypothesis: Dark Matter Redistribution**

The hierarchical galaxy formation process proceeds as is currently understood, until the stage where the early galaxies form. Based on recent observations, these early galaxies have grown in mass rapidly perhaps as predicted by N-body simulations taking into account cold gas accretion in massive halos (Kang *et al* 2010). A significant number of these early massive galaxies grow to masses comparable to today's galaxies by about 3 BY (Glazebrook *et al* 2004), and are much more compact (Trujillo *et al* 2007; Buitrago *et al* 2008; vanDokkum *et al* 2009). It is at this time that the central black holes (BH), so far found in virtually every galaxy including those with no significant bulge (Satyapal *et al* 2008), become active galactic nuclei (AGN). AGN activity during this time is $10^4$ times larger than today (Whittle 2008, Chapters 20 and 21). During the period of AGN activity, the central BH grows and gains rotation, as mass accretes into it. A BH of maximum rotation has 0.29 of its mass contributed by rotational energy (Harrison 2003, Chapter 13). Roger Penrose showed, theoretically, that mass entering a BH's ergosphere can emerge with higher energy (as a component falls into the BH and another component escapes with more energy than the in-falling mass). The energy gained by the emerging mass is at the expense of the rotational energy of the BH. In principle, the Penrose process can extract all the rotational energy of the BH (Harrison 2003, Chapter 13). In order for the Penrose process to apply there must be a single object whose momentum is conserved at the time of splitting into the two components. Thus, the dark matter either decomposes into two components, as has been previously suggested by Grib and Pavlov (2009); or alternatively, forms a composite object, with either visible matter or dark matter, that breaks into two components. The two components must consist of a heavy component and a much lighter one (at least $10^{-5}$ lighter); where the lighter component falls into the BH and the heavier component escapes.

Let us assume that the dark matter near the cusp of the halo of the compact object accelerates via the Penrose effect, and achieves a change in velocity, or a $\Delta V$, as is commonly used in orbital mechanics. If the $\Delta V$ is sufficient to have the particle achieve a velocity in the order of the escape velocity, then the dark matter particle will go into a highly elliptical orbit. If enough dark matter particles receive a $\Delta V$, then the dark matter will be redistributed from a compact space into elliptical orbits, yielding a very different and more spread out dark matter distribution. The dark matter must only interact gravitationally and have inertia (i.e. be 'dark'), for the above redistribution to occur.

In section 3, I develop, at a high level, the dark matter redistribution hypothesis to explore its impact on galaxies and their formation history.

## 3. Mathematical Development

Table 1 contains all the key parameters used in subsequent calculations.

Table 1 - Galaxy Parameters.

|  | HSB galaxy today | Compact Early Galaxy | LSB galaxy today |
|---|---|---|---|
| Mass | $10^{42}$ kg (i.e. $5 \times 10^{11}$ $M_O$) | $10^{42}$ kg | 0.2 HSB mass |
| M/L | 10 | 10 | 30 |
| Extent of most of Dark Matter | 30 kps | | |
| Effective radius | ~3-4 kps | ~0.8 kps[B] | |
| Maximum rotation velocity | ~250 km/sec | | ~120 km/sec |
| Bulge Mass | 1/2 the visible mass | | |
| Black Hole mass | 0.0012 Bulge Mass[A] | | |

HSB = High Surface Brightness
LSB = Low Surface Brightness
$M_O$ = Solar mass
M/L = mass to light ratio of the galaxy
References. (A) McLure 2002; (B) vanDokkum 2009

*3.1 Dark Matter Redistribution*

Assume that a significant portion of the dark matter is in orbits within the early galaxy's observed size. As dark matter enters the ergosphere of the central BH, some of it emerges with increased energy. By the time the dark matter returns to the original confinement radius it has acquired a change in velocity, $\Delta V$. The $\Delta V$ causes the dark matter to enter an elliptical orbit. For a circular orbit of radius $r_1$, a $\Delta V$ (along a circular orbit) produces an elliptical orbit; with perigee $r_1$ and apogee $r_2$ related to $\Delta V$ (Meriam 1975) by

$$\Delta V = \sqrt{\frac{G M_e}{r_1}} \left( \sqrt{\frac{2 r_2}{r_1 + r_2}} - 1 \right), \qquad (1)$$

where $M_e$ is the mass enclosed by the original circular orbit.

The *ΔV* for escape is

$$\Delta V_{esc} = \sqrt{\frac{G M_e}{r_1}} \left(\sqrt{2} - 1\right), \qquad (2)$$

assuming the entire mass is enclosed within $r_1$. As *ΔV* approaches the magnitude of the *ΔV* for escape, $r_2$ becomes many times $r_1$, and the resulting orbit is highly elliptical. In our case, $r_1$ is of the order of the size of the initial structure, and $r_2$ is perhaps 10 to 30 to 100 times larger. For a general case, the *ΔV* values are larger (by up to a factor of 2.4) than the above results obtained for the circular orbit case.

Assume, for now, that some dark matter receives a *ΔV*, as above, and enters into a highly elliptical orbit, say with $r_2 \sim 30$ kps. In an elliptical orbit with central force, angular momentum is conserved (Keppler's second law) and therefore,

$$\frac{d\theta}{dt} r^2 = h, \qquad (3)$$

Where $\theta$ is the angle along the plane of the orbit, $r$ is the radius of the orbit, and $h$ is a constant. For any small angle, $d\theta$, the mass will dwell in that angle for time, $dt$, where

$$dt = \frac{d\theta}{h} r^2 . \qquad (4)$$

Assume that many particles enter into the same orbit around the center of the original structure. The amount of mass at any radius $r$ will be proportional to the time spent by mass at that radius, $dt$. Integrating (4), to find the time required by any particle to achieve a certain radius $r$ from $r_1$, will lead to a result proportional to the enclosed mass within that radius, from all the orbiting particles. Enclosed mass will be zero for $r < r_1$ and be equal to the total mass of orbiting particles for $r > r_2$. Figure 1 shows the result of numerical integration of (4) for an ellipse with eccentricity of 0.93; displayed against unitary mass and unitary maximal radius. Results for other eccentricities in the 0.6 to 1 range are similar. Also shown in figure 1, is a mass proportional to $r^2$ curve. The fit to $r^2$ is within 10 per cent for radii from about 0.4 to 0.97; roughly the portion of rotation curves dominated by dark matter.

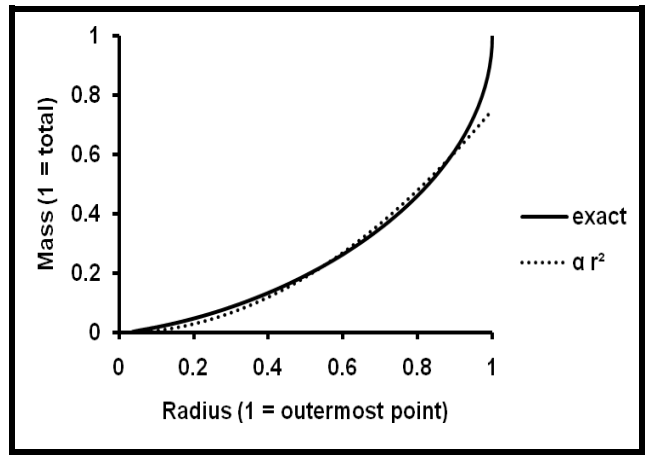

Figure 1 - Enclosed mass vs. radius.

The process described above means that some of the original, tightly cusped shaped, profile of dark matter has developed into an approximately constant surface density profile (mass proportional to $r^2$ approximation) seen along the galaxy disk.

Assume that the majority of the mass undergoes the process described above. **What happens to the last dark matter particle to receive a $\Delta V$?** By that time, the matter distribution of the galaxy is leading to the typically observed rotation curves (as demonstrated in section 3.3). These rotation curves imply a range of potentials (Struck 2006), most commonly; either a logarithmic potential (for typical HSB galaxies), or an $r^1$ potential (for rising portions of typical LSB galaxies), versus the $r^{-1}$ potential experienced from a concentrated central mass used in the above analysis. For a range of potentials, including both the logarithmic and the $r^1$ potentials, it has been derived (Struck 2006) that the orbits are approximately elliptical in nature. Therefore, the result derived from (4) holds. The different potentials demand different $\Delta V$'s in order to attain a particular maximal radius orbit. For a logarithmic potential, calculations (for the typical HSB galaxy in table 1) demonstrate that the last dark matter particle requires about 15 per cent of the extra energy, required by the first dark matter particle, to reach the same 30 kps maximal orbital radius. By the time the last dark matter particle achieves its orbit, the orbit of the first dark matter particle has changed, as dark matter has been redistributed. The orbit remains nearly elliptical in shape (Struck 2006). Calculations demonstrate that the orbital parameters ($r_1$, $r_2$) adjust and increase (since the mass enclosed by their orbit is decreasing due to redistribution; see section 3.2), depending on their size, compared to the same parameters of the remaining dark matter particle orbits. In fact, the orbital parameters of other particles that have not received a $\Delta V$ also adjust in the same manner.

The picture that emerges is, that dark matter particles receive a $\Delta V$ (different for each particle depending on its entry into the ergosphere, and possibly varying with time depending on AGN activity), and enter various elliptical orbits. For a fixed $\Delta V$, each succeeding particle enters an orbit with slightly higher maximal radius, as it has to overcome less potential energy. As the mass distribution changes (shifting from the inner section of the galaxy to the outer section), the orbits of the particles, already in elliptical orbits (either those that have received a $\Delta V$ earlier or those that have not), adjust but remain approximately elliptical in shape. The resulting parameters of the orbits of dark matter particles vary; however, for any group of dark matter particles with similar orbital parameters, their mass enclosed within a radius $r$ is as shown in figure 1. As will be demonstrated in section 3.3 below, mass distributions consisting of different amounts of dark matter particles in orbits, with different maximal and minimal radii, have a relatively small effect on the overall shape of rotation curves.

As long as a significant portion of the dark matter is placed into highly elliptical orbits causing central density of dark matter to drop, the rest of the dark matter (and visible matter see section 3.2) will also be redistributed. The impact from the redistribution of dark matter will be significant. Recent detailed studies reveal that the dark matter halos around spirals have densities of up to one order of magnitude lower than the $\Lambda$ CDM predictions for the inner regions. At about 2 to 3 times the optical radius, they have densities higher, by a factor of 2 to 4, than the corresponding NFW predictions (Salucci and Martins 2009).

Finally, since the dark matter inside the cusp receives a $\Delta V$ in the plane of rotation of the BH, it enters into stable elliptical orbits (as has been assumed above) that favor low inclinations to the plane of rotation of the BH. The dark matter that has not received a $\Delta V$ still occupies space perpendicular to that plane.

**Is there enough energy available to achieve the scenarios described above?** The mass energy available is some percentage (less than 29 per cent since most galaxy BHs are close to maximum rotation (Fanidakis *et al* 2009)) of the BH mass. It has been found that BH mass correlates strongly with bulge mass (for galaxies with bulges, BH mass = 0.0012 bulge mass (McLure 2002)), and assuming that bulge mass is approximately ½ the visible mass, we obtain that the mass energy available from the BH is

$$\beta \; 0.0012 \; \frac{1}{M/L} \; \frac{1}{2} \; M_G \; c^2, \qquad (5)$$

where $\beta$ is the portion of BH mass used, $M/L$ is the mass to light ratio of the galaxy, $M_G$ is the mass (visible and dark) of the galaxy, and $c$ is the speed of light. To provide a $\Delta V$ to all the dark matter (5) must equal (assuming large mass to light, $M_G \sim$ mass of dark matter)

$$\frac{1}{2} M_G \left( (V + \Delta V)^2 - V^2 \right), \tag{6}$$

where $V$ is the original orbital velocity of dark matter particles. Using (2), with the estimate that due to the decreasing potential (on average) particles require 25 per cent or less of the escape energy to achieve dark matter redistribution, (6) approximately equals

$$\frac{1}{2} M_G \left( \frac{G M_G}{r_1} \right) 0.25 . \tag{7}$$

For our HSB galaxy, $\beta$ is about 5 per cent, in the order of magnitude of the theoretical 29 per cent limit. This calculation underestimates the value of $\beta$ because the original mass is not in a circular orbit. However, it over estimates the value of $\beta$ because less than all the dark matter need receive a $\Delta V$ to achieve redistribution of all the dark matter, and only about half of the mass is initially contained in the effective radius. Notice that only the $M/L$ ratio and the $M_G/r_1$ ratio enter the calculation. As long as initial compact structures have a total mass to size relation, and a reasonable range of $M/L$ ratios, the calculation above leads to a similar result for other galaxies.

Finally, it should be noted that nothing is stopping a portion of the original dark matter in the structure from obtaining enough $\Delta V$ to escape the galaxy, resulting in a remaining structure which is lighter in dark matter; however, still exhibiting a dark matter mass distribution described above.

**What is the duration of dark matter redistribution?** Redistribution of dark matter is possible only after the galactic nucleus has become active, causing the central BH to gain rotational energy. Moreover, since the energy requirements for redistribution of dark matter are of the same order as the rotational energy of the BH, redistribution must span the full AGN period, typically thought to be about 1 BY. However, some redistribution could continue for some time past the AGN period, as long as some accretion is occurring. Half the orbital period of the dark matter is a fraction of a BY. Therefore, most of the redistribution must happen in a period of the order of ~ 1 BY.

**Can the black hole capture enough dark matter during this period to achieve the redistribution?** Calculations following the method employed by Grib and Pavlov (2009) show that that the BH can capture the dark matter within the early galaxy over a period of 1 BY. The BH dark matter capture rate is given by

$$\sigma_c \upsilon_o \rho , \tag{8}$$

where $\sigma_c$ is the BH capture cross section, $\upsilon_o$ is the average dark matter particle velocity (~$10^6$ m/sec), and $\rho$ is the dark matter density (averaged over several orders of magnitude times the Schwarzschild radius of the BH, i.e. ~ a fraction of a parsec).

The capture cross section for non relativistic particles by a Schwarzschild BH is given by (Zel'Dovich and Novikov 1971)

$$\sigma_c = 4\pi \left( \frac{c}{\upsilon_o} \right)^2 r_g^2 , \tag{9}$$

where $r_g$ is the horizon radius of the BH (i.e. ~$3 \times 10^{-5}$ parsec). The capture cross section for a rotating BH is of the same order (Zel'Dovich and Novikov 1971). Using an NFW profile for the dark matter in our initial galaxy yields a $\rho \sim 3 \times 10^{-11}$ kg/m$^3$ over the inner fraction of a parsec of the galaxy; which in turn leads to a capture rate of ~ $3 \times 10^{25}$ kg/sec. This capture rate times a BY, is approximately equal to the total dark matter in the early galaxy. However, for the BH to sustain this capture rate for a BY, the dark matter must be constantly brought to within less than a parsec of the center compact galaxy from its full extent (i.e. ~ few kps). The dark matter particles will continuously pass close to the center of the galaxy if they are primarily in box orbits (which have a period of a small fraction of BY), as will be the case if the initial dark matter structure is triaxial as many simulations have shown (Frenk *et al* 1988; Bellovary *et al* 2008).

*3.2 Impact on Visible Matter*

Consider visible matter in approximately circular orbits in the plane of rotation of the BH. The forces must balance, therefore,

$$V^2 = \frac{G M_e}{r}, \qquad (10)$$

where $V$ is the rotational velocity of the visible matter in the orbit, $r$ is its orbital radius, and $M_e$ is the mass enclosed by the orbit. As dark matter is redistributed, $M_e$ drops. Forces are all radial to first order, therefore, angular momentum is conserved, and

$$rV = \text{constant.} \qquad (11)$$

Combining (10) and (11) we obtain

$$r \alpha \frac{1}{M_e} \qquad (12a)$$

and

$$V \alpha M_e. \qquad (12b)$$

As $M_e$ drops, visible matter orbital radius increases, and visible matter orbital velocity decreases.

Assume, at the outset, that our galaxy has approximately 80 per cent of the dark matter (a large fraction is chosen to estimate largest feasible growth factor), and 100 per cent of the visible matter enclosed at the outer observed radius (approximately 2 to 3 times the 0.8 kps in table 1). After redistribution, the visible matter continues to be enclosed by the outer observed radius (of stars); however, the same radius encloses a smaller portion of the dark matter. For a typical HSB galaxy today, the enclosed mass of dark matter, at the radius of the outer stars (say the sun's location around 8 kps), is approximately one visible mass. Therefore, for an *M/L* of 10, the enclosed mass is reduced from ~8 times the visible mass to ~2 times the visible mass:

- $M_e$ decreases by a factor ~4 (i.e. the M/L within the extent of the visible mass decreases from ~8 to ~2),
- $r$ increases by a factor of ~4, and
- $V$ decreases by a factor of ~4.

The extent of visible matter grows, and its rotational velocity drops, both effects occurring at the same time as the dark matter is redistributed. The enclosed dark matter mass after growth is approximately proportional to *M/L* times $r^2$, limiting the growth of structures as the M/L ratio increases. Therefore, repeating the calculation for galaxies with larger *M/L* yields only a slightly larger growth factor in the extent of the visible mass. The calculation above is dependent on the chosen radius, since the enclosed dark matter varies with radius differently than the enclosed visible matter.

As was discussed in section 3.1 the redistribution of dark matter favors the BH's plane of rotation. Thus, the simplified analysis above is correct only for stars in orbits in the plane of rotation of the BH. Calculations for orbits off the plane of BH rotation are more complex since angular momentum is not conserved. These orbits also grow in size and in addition experience torques towards the BH's plane of rotation.

The effect described above for circular orbits also applies to elliptical orbits (although the calculations are more complex) and results in an increase in orbital radius (and in particular in $r_1$ and $r_2$) of the orbits of dark matter particles.

*3.3 Impact on Galaxy Rotation Curves*
The rotation velocity of objects in a spiral galaxy is dependent on the enclosed mass at any given radius. The enclosed mass is comprised of the mass of the stars, the mass of the gas, and the mass of the dark matter.

Star mass can have two distributions; a central dense bulge and a more spread out disk. HI gas mass distribution approximately follows dark matter distribution (Hoekstra *et al* 2001). Other gases are usually immaterial to the shape of the rotation curve.

Since we are interested in exploring the impact of the dark matter redistribution, we assume a simple visible matter distribution; a constant density until a particular radius, and zero beyond that radius. We ignore the distribution of the mass of the HI gas since it approximately follows the dark matter distribution. The dark matter mass distribution is complex, as discussed in section 3.1. We assume each dark matter particle receives a change in energy that is a fraction of the change in energy required for escape. Each particle enters into an approximately elliptical orbit with low inclination with respect to the to the galaxy's BH plane of rotation. Subsequent particles enter orbits with progressively larger maximal radii. The orbits of particles that have already received a *ΔV,* and those that have not, grow in extent. Approximate calculations reveal that at the end of dark matter redistribution the elliptical orbits occupy a small range of $r_1$'s (typically the first few kps for our HSB galaxy) and a range or $r_2$ (typically spanning the outer half of the galactic radius (i.e. ~ 15 to 30 kps)). In the very core of the galaxy, density is zero. In the inner part (first few kps), the staggering of $r_1$'s leads to an approximately linear rise in dark matter surface density (i.e. constant density); in the middle section (up to ~15 kps), the dark matter surface density transitions to approximately constant. In the outer half of the galactic region (~15 - 30 kps), the dark matter surface density falls as mass from some orbits is completely enclosed, while mass from other orbits is still growing (approximately proportional to $r^2$).

The visible mass rotational velocity component increases linearly, followed by the $r^{-1/2}$ keplerian decline, as derived from (10). The dark matter rotational velocity is; at first zero; subsequently increasing linearly (i.e. (10) with constant density); next, increasing approximately as $r^{1/2}$ (i.e. (10) with constant surface density); and finally, approximately flat (i.e. (10) with decreasing surface density).

The visible matter and dark matter components result in the galaxy rotation curve. For a typical HSB galaxy, the rotation curve is as shown in figure 2a.

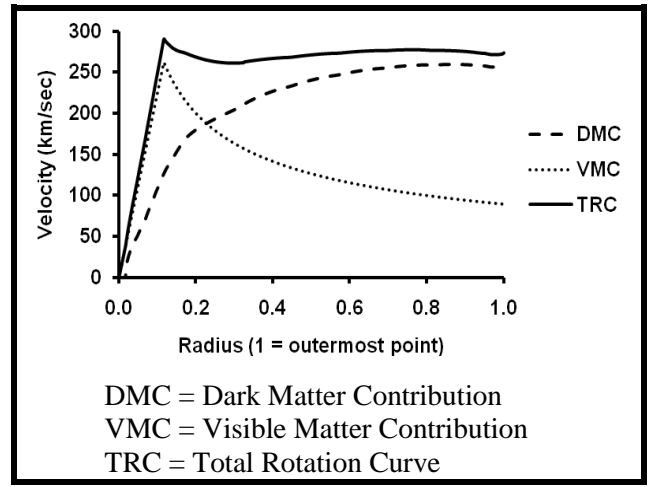

DMC = Dark Matter Contribution
VMC = Visible Matter Contribution
TRC = Total Rotation Curve

Figure 2a – High Surface Brightness (HSB) galaxy rotation curve.

In the transition region, from visible matter dominance to dark matter dominance, the two rotation curve components result in a roughly flat rotation curve; as the dark matter component increases approximately proportional to $r^{1/2}$, and the visible matter component decreases proportional to $r^{-1/2}$. This assumes a constant HSB galaxy example density of visible matter up to a radius of ~12 per cent of the maximal extent of the dark matter.

Decreasing the overall mass, increasing the *M/L* ratio, and spreading the visible matter over a larger radius, one obtains the typical LSB galaxy rotation curve shown in figure 2b. The curve climbs quickly; at first from the combined rising visible matter and dark matter components, subsequently more slowly as the visible matter component declines, and finally is approximately flat.

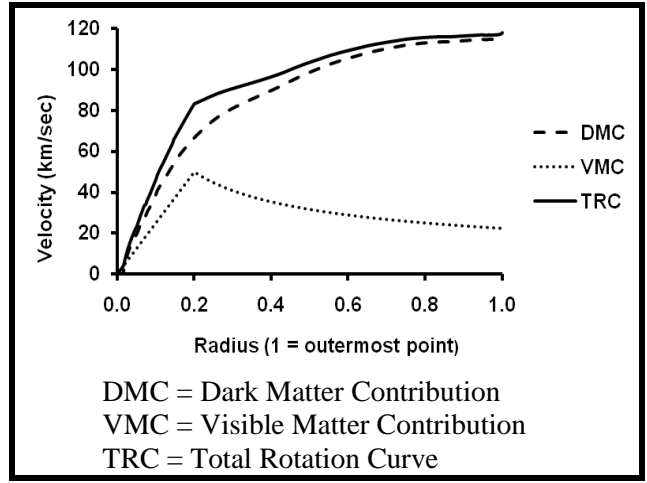

DMC = Dark Matter Contribution
VMC = Visible Matter Contribution
TRC = Total Rotation Curve

Figure 2b – Low Surface Brightness (LSB) galaxy rotation curve.

Thus, the redistribution of dark matter and associated visible matter spread (ignoring the significant structure at small radius for the bulge component) naturally leads to the typical rotation curves. In fact, if the dark matter goes into smaller elliptical orbits, the enclosed mass at any radius is larger, and the visible matter expands to a smaller radius, and vice versa. Using (4), (10), (12a) and (12b) one can derive a relation between the ratio of the extent of dark matter to the extent of visible matter, and the expansion ratio of visible matter for any given *M/L*. In other words, dark matter and visible matter extents are coupled and dependent on the initial compact structure radius and its *M/L* ratio. The *M/L* ratio dependence is weak, and is immaterial for *M/L*>20.

Approximate simulations of the redistribution of dark matter, in different proportions to elliptical orbits with different radii, reveal that the shape of rotation curves are not materially affected. These simulations effectively emulate AGN activity periods of different duration, redistributing dark matter with slightly different *ΔVs*. The panels in figure 3 show rotation curves for different dark matter redistributions, in HSB galaxies and LSB galaxies. The first row repeats Figs. 2a and 2b for comparison purposes. The second row shows more dark matter redistributed to the small radii and less to the larger radii. The third row shows more dark matter redistributed to the middle radii and less to the other radii. The fourth row shows more dark matter redistributed to the larger radii and less to the smaller radii. Finer structure develops without a significant effect on the overall shape of the rotation curves.

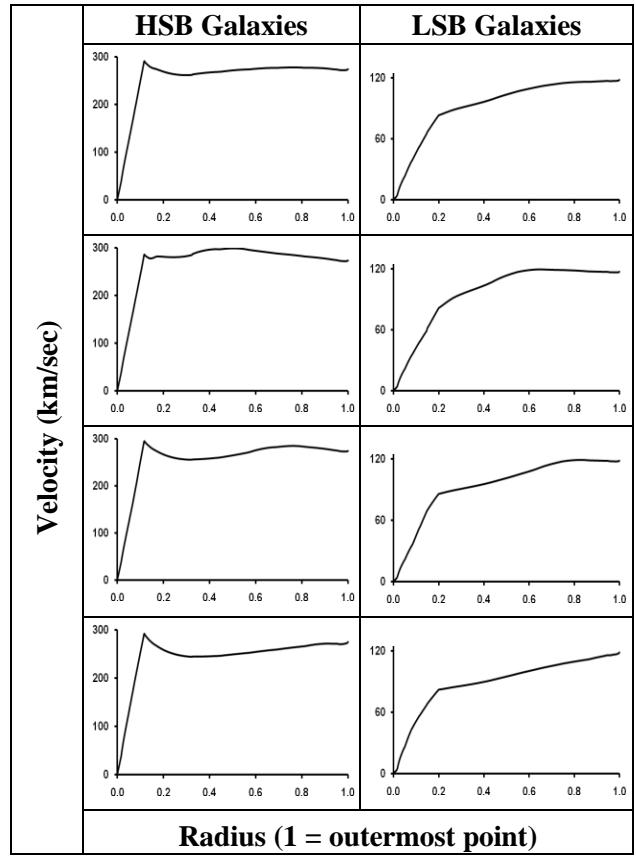

Figure 3 – Rotation Curves for redistribution of dark matter in different proportions to different maximal radii.

*3.4 Impact on Galaxy Evolution History*
The results presented above lead to the following description of galaxy evolution. The galaxies form to a large mass early on. While the majority of galactic nuclei are active (2 to 5 BY after the Big Bang), galaxies redistribute their dark matter. During the same period, the visible matter unfolds from compact orbits to larger orbits (a few times larger, depending on *M/L* ratios). When the redistribution of matter is complete (5 to 7 BY after the Big Bang, and in a timeframe of ~1 BY for any particular galaxy), the galaxies have transformed from compact structures into highly organized spiral structures. These galaxies have acquired correlated dark matter and visible matter components; have grown in size but not mass (although the *M/L* ratio in the visible part of the galaxy has decreased significantly); and, have acquired 'flat' rotation curves. They appear similar to those observed nearby (Sparke and Gallagher 2007, Chapters 8 and 9; Williams *et al* 2010). Thereafter, for the next ~7 to 9 BY, if they are in a crowded space, then they undergo further development by collisions and mergers (including further growth in size and mass, and the formation of elliptical galaxies from equal mergers); or, if they are isolated, then they evolve on their own.

**4. Discussion**
The modification of the hierarchical galaxy formation theory described above affects the evolutionary history of galaxies and their resulting visible matter and dark matter distribution.
   The distribution of matter in galaxies is examined by studying galaxy rotation curves. As shown in section 3.3, the redistribution of dark matter and associated visible matter spread naturally leads to typical observed rotation curves. However, unlike the standard hierarchical galaxy formation theory, which predicts cusped shaped dark matter halos and uncorrelated galaxy parameters, the modification to the theory predicts results that more closely agree with key observations. As further shown in section 3.3 there is a relation between the ratio of the extent of dark matter to the extent of visible matter, and the expansion ratio of visible matter for any given *M/L*.  This relation means: (1) that dark matter and visible matter extents are coupled and dependent on the initial compact structure radius and *M/L* ratio, and (2) that the rotation curves have a radial scale matched to the maximum dark matter extent.  These results imply that the shapes of the rotation curves are extremely resilient and consistent, and explain the 'halo-disk conspiracy' and the correlation between basic galaxy dynamic parameters (light radii and dynamical mass).  In addition, the redistribution of dark matter during the galaxies development naturally transports dark matter concentrated near the center of the structure (the cusped shaped halo originally developed via hierarchical merging) to the outer regions of the galaxy, producing a core shaped dark matter halo.

As discussed in the Introduction, various models for the evolution of galaxy size and mass with time have been proposed and tested (Hopkins *et al* 2009; vanDokkum *et al* 2009).   These models fail to explain some key aspects of observations of equal mass galaxies at different look back times made by Buitrago, Trujillo and their co-authors.  The modification of the hierarchical galaxy formation theory, which leads to the galaxy evolutionary history described in section 3.4, provides an explanation.

Buitrago, Trujillo and their co-authors plot the size of equal mass spheroid-like objects and disk-like objects from *z* ~ 3 to the present (Trujillo *et al* 2007; Buitrago *et al* 2008).  One can look at the plots and envisage evolution for both types of structures separately from *z* ~ 3 to the present (i.e. both disks and spheroid-like structures become larger with time).  Alternatively, applying the modification described herein, one can read the results differently.  The equal mass spheroid-like structures evolve from mergers of earlier structures (perhaps from violent mergers of gas rich discs as proposed by  Ricciardelli *et al* (2010)) and reach their given mass at different *z* ranging from ~3 to ~1.7.  The plot of the size of spheroid-like objects for *z* greater than 1.7 is consistent with them being approximately equal in size (Buitrago *et al* 2008).  As AGN activity builds, these spheroid-like objects redistribute their matter over one BY and become disk-like.  The plots show an effective radius growth factor between the spheroid-like and disk-like structures  (allowing for 1 BY of elapsed time) of ~2.6, consistent with the dark matter redistribution modification.  The newly formed disk-like structures continue evolving to the present in one of two ways:  minor mergers (with their associated further growth in structure size and mass (Hopkins *et al* 2009)) and development of spiral arms and bars (as shown by N-body simulations) leading to today's spiral galaxies; and, major mergers leading to today's elliptical galaxies.

The growth in effective radius by a factor of ~2.6 should result in a decrease in velocity dispersion by approximately the same factor, per section 3.2.  This result is consistent with velocity dispersion measurements published by vanDokkum *et al* (2009) that show a decrease in velocity dispersion, for a galaxy in the same mass range, by a factor of 2.55 from *z* = 2.186 to the present.  However, the result is inconsistent with measurements published by Cenarro and Trujillo (2009) that show a much smaller change in velocity dispersion.  Their data is from z = 1.6 to present and may represent the evolution due to mergers only (with their associated small change in velocity dispersion (Hopkins *et al* 2009)) versus the major evolution due to redistribution of matter (with its associated larger change in velocity dispersion) that would have happened at larger z (i.e. z~2).

The proposed evolutionary picture (see section 3.4) also predicts that there should be virtually no remnants of the original compact structures today, as has been observed (Trujillo *et al* 2009); as opposed to the prediction of a non-negligible fraction (1 to 10 per cent) of remnants from some hierarchical merging scenarios (Trujillo *et al* 2009).

The following issues, that bear on the correctness of the modification, remain to be confirmed when more data becomes available.

First, the current data shows approximately equal numbers of spheroid and disk-like structures at high *z* (2.5 to 3.0) (Buitrago *et al* 2008).  Although AGN activity is already high by that time, allowing for matter redistribution and production of disk-like structures from spheroids at an earlier period, it would be more comforting if the number of spheroid-like objects was larger than disk-like objects; as would have to be the case at z higher than 3.5 (before AGN activity peaked).  Alternatively, some of the disk-like objects could evolve directly from disk-like predecessors in the early universe without merging to become spheroid-like objects (i.e. triaxial) and undergoing dark matter redistribution; however, one would not then expect them to have the correlated dark and visible matter components and associated rotation curves discussed herein.  The current data also shows spheroid objects at z = 1; however, these can be explained as the result of mergers of lower mass disks that formed earlier.

Second, the proposed modification predicts a fast early evolution in size (followed by a gradual evolution caused by mergers) as opposed to other explanations, that do not modify hierarchical galaxy formation theory, and that predict a more gradual evolution (Hopkins *et al* 2009). The latest galaxy growth studies by vanDokkum *et al* (2010) remain inconclusive as to how fast the size evolution is for the period prior to z=1.5. However, there is evidence emerging for a rapid evolution in galaxy size. Mancini *et al* (2010) investigated 12 massive quiecent early type galaxies at $z \sim 1.5$ and found that 9 of those galaxies follow the same size - mass relation of local ellipticals. Although the authors question earlier size measurements of high z compact galaxies, Szomoru *et al* (2010) have reconfirmed that a massively compact galaxy at $z = 1.91$ lies significantly off the local size - mass relation. Thus, these results indicate that at least some compact galaxies at high z have evolved in size rapidly; by $z \sim 1.5$.

Finally, the proposed modification predicts a large change in velocity dispersion vs the late dry merger model (Hopkins *et al* 2009) which predicts a smaller change in velocity dispersion; and once again the observational evidence remains inconcluisve (Cenarro and Trujillo 2009; vanDokkum *et al* 2009).

## 5. Conclusions

The presently favored hierarchical galaxy formation theory cannot completely explain key observations, related to both galaxy evolution and current dark matter distribution in galaxies. Modifying the theory to include a stage, during the AGN phase of the galaxy's life, where a significant portion of the dark matter is placed into highly elliptical orbits, brings theory and observation into closer alignment. The energy required to place dark matter into elliptical orbits is available from the rotational energy of the galaxy's central BH. By placing a significant portion of the dark matter into elliptical orbits, the resulting mass distribution leads to the characteristic galactic rotation curves. Finally, the redistribution of dark matter causes the extent of visible matter to grow; helping to explain observations that show early galaxies were much more compact.

Further work is required to test the dark matter redistribution hypothesis by comparing comprehensive and detailed numerical modeling results of the ideas presented here to current observations; and, as they become available, to improved measurements of the velocity dispersion in early galaxies and higher fidelity data on the rate of evolution of galaxy size.

**Acknowledgement**
The author would like to thank Dr. Ignacio Trujillo for his insightful questions and suggestions.